\documentclass[aps,prl,reprint]{revtex4-1}
\usepackage{graphicx}  
\usepackage{epstopdf}
\usepackage{bm}        
\usepackage{amssymb} 
\usepackage{ulem}
\usepackage{color}

\begin{document}
\title{Disclinations, e-cones, and their interactions in extensible sheets}
\author{Julien Chopin}
\email{julien.chopin@espci.fr}
\affiliation{Gulliver UMR 7083, CNRS,  ESPCI ParisTech,
PSL  Research  University,  F-75005  Paris,  France}
\affiliation{Sorbonne Universit{\'e}s, UPMC Univ Paris 06, CNRS, UMR 7190, Institut Jean Le Rond d'Alembert, F-75005 Paris, France}
\author{Arshad Kudrolli}
\email{akudrolli@clarku.edu}
\affiliation{Department of Physics, Clark University,  Worcester, MA 01610, USA}

\begin{abstract}
We investigate the nucleation, growth, and spatial organization of topological defects with a ribbon shaped elastic sheet which is stretched and twisted. Singularities are found to spontaneously arrange in a triangular lattice in the form of vertices connected by stretched ridges that result in a self-rigidified structure. The vertices are shown to be negative disclinations or e-cones which occur in sheets with negative Gaussian curvature, in contrast with d-cones in sheets with zero-Gaussian curvature. We find the growth of the wrinkled width of the ribbon to be consistent with a far-from-threshold approach assuming a compression-free base state. The system is found to show a transition from a regime where the wavelength is given by the ribbon geometry, to where it is given by its elasticity as a function of the ratio of the applied tension to the elastic modulus and cross-sectional area of the ribbon.
\end{abstract}
\maketitle

Localized defects in the form of disclinations, grain boundaries, voids, and inclusions that mark an otherwise featureless solid are important to understanding and designing the macroscopic properties of materials~\cite{Chaikin2000}. Topological defects such as disclinations and dislocations are known to control
the morphology and mechanical properties of thin flexible sheets and membranes~\cite{Nelson2002}. For example, a disclination appears in a thin disk shaped sheet when the metric of a surface is modified by adding a wedge.
Such a point-like defect induces in-plane stresses that can be  alleviated by out-of-plane deformations~\cite{Seung1988}. Disclinations can be positive or negative, depending on the ``charge" associated with Gaussian curvature around the defect. Oppositely signed disclinations can pair up resulting in what is called a dislocation with zero net Gaussian charge~\cite{Nelson2002,Wang2009,Guven2013}. Thus, a sheet with a dislocation is isometric to a plane outside the core of the defect. In the context of elastic sheets, dislocation and negative disclinations are usually called d-cones and e-cones, respectively~\cite{BenAmar1997,Muller2008}.  Recent studies suggest a deep connection between topological defects, such as disclinations and disclocations in crystaline membrane, and e-cones and d-cones in amorphous membranes based on their Gaussian charge~\cite{Santangelo}.

\begin{figure}
\centering
\includegraphics[width = 8cm]{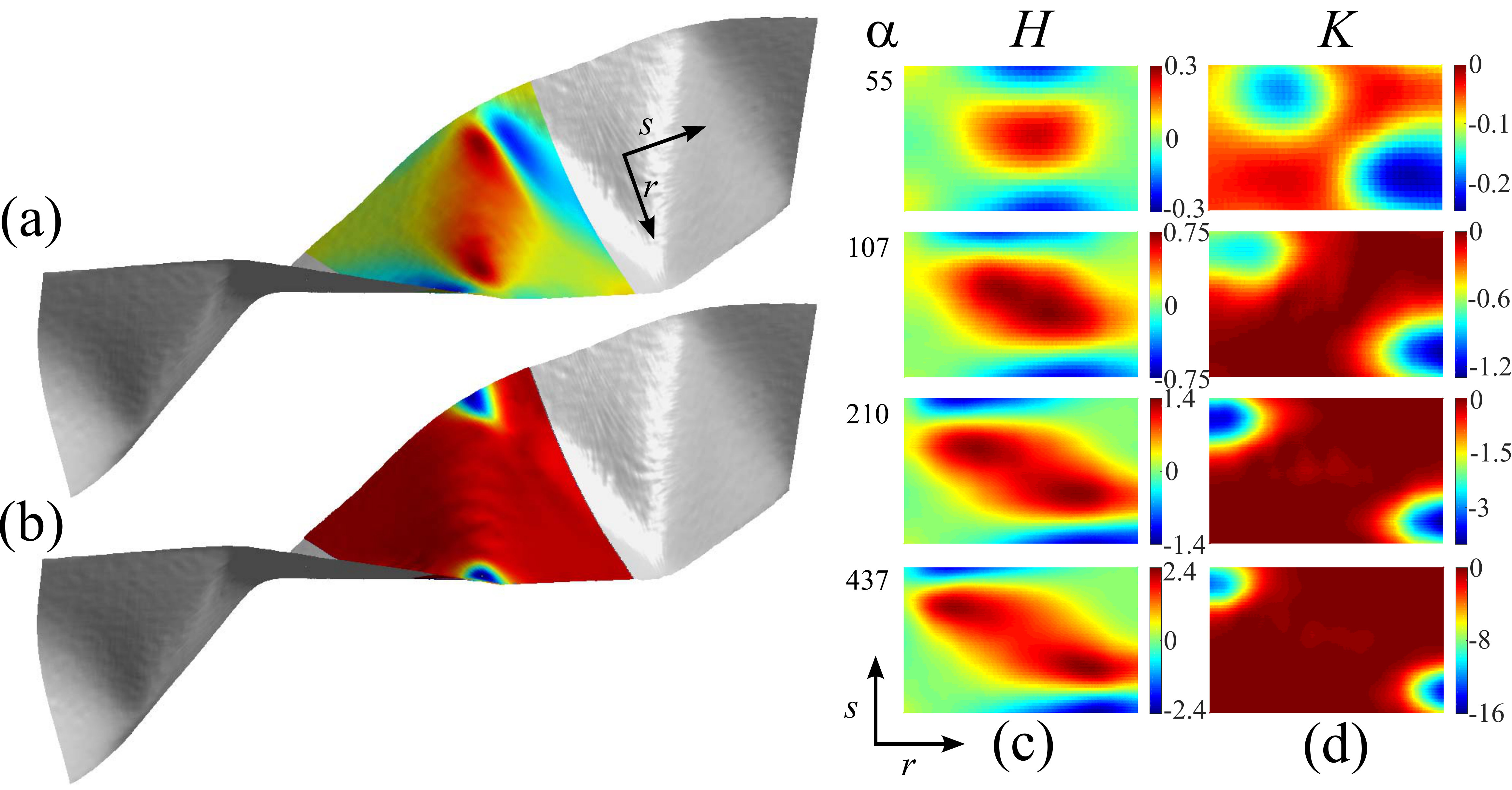}
\caption{A 3D reconstruction of a twisted ribbon (Mylar, $t/W = 8.5 \times 10^{-3}$, $T = 1.2 \times 10^{-3}$ and $\alpha = 437$) with measured (a) Gaussian curvature $K(r,s)$, and (b) mean curvature $H(r,s)$ superimposed on a ridge. Map of (c) $H(r,s)$  and (d) $K(r,s)$ as a function of increasing $\alpha$. Here we show that a symmetry breaking and a triangular lattice develop with non-zero Gaussian curvature. The lattice spacing appears almost constant with $\alpha$.}
\label{fig1}
\end{figure}
Considerable experimental and theoretical challenges exist to identify and model the emergence of such defects and their dynamics under external forcing. Isolated defects in infinite sheets have been well studied~\cite{BenAmar1997,Chaieb1998,Cerda1998,Chaieb1999,Cerda1999,Muller2008,Guven2013,Efrati2015}. Outside the core of the defect, the deformations are assumed to be inextensible, i.e. stress free. While this has been shown to lead to a reasonable description of the overall shape of the surface~\cite{Cerda1998}, the inner structure of the defect and its interaction with other defects and surface edges
are still not well understood~\cite{Witten2007}. 
Because interacting defects can be commonly noted as in crumpled paper~\cite{Lobkosvky1995,Blair2005,Witten2007,Aharoni2010} and indented shells~\cite{Pauchard1998,Boudaoud2000,Hamm2004}, a detailed geometrical characterization of defect interactions in sheets under well defined loading and boundary conditions is still needed to build a deeper understanding of macroscopic properties of sheets undergoing large displacement. 

Because of its rich phase diagram and well-defined boundary conditions, the twisted ribbon configuration has been proposed as a model system to understand the nonlinear and singular behavior of elastic sheets~\cite{Chopin2013,Chopin2015}. A particularly interesting aspect of the system is the spontaneous emergence of ridges and point-like defects organized in a triangular lattice that have been shown to form at small tension~\cite{Korte2011}. This simplified geometry develops rigidity due to the formation of ridges and allows investigation of the formation of interacting singularities under well-defined loading conditions.   Over the last decades, various theoretical approaches have been proposed to model twisted ribbons including anisotropic rod-like theory~\cite{Goriely2001,Korte2011,Dias2015} and nonlinear plate and shell theory~\cite{Green1937,Mansfield2005,Coman2008,Mockensturm2001,Efrati2009,Davidovitch2011}. However, none of these approaches have captured or predicted the spontaneous emergence of the ridges and their fine structure with the exception of a recent study which has sought to describe some aspects of the extension of a creased sheet with far from threshold analysis~\cite{Chopin2015}. 

Here, we address the nucleation of topological defects in elastic sheets, and examine their growth and detailed structure using micro-focus x-ray computed tomography. We find that the structure of the observed vertex singularities  are different from d-cones which are often assumed while viewing such defects in thin sheets. In fact, we find these defects correspond to negatively signed interacting disclinations, i.e. interacting e-cones. We show that triangles emerge even when stretching is applied, in contrast with analysis with inextensible sheets by Korte, {\it et al.}~\cite{Korte2011}. We measure the growth of the width of the wrinkles and show that it is consistent with a far-from threshold approach assuming a compression free base state~\cite{Chopin2015}. We find that the wavelength of the wrinkles changes from being given by ribbon geometry to its elasticity as the tension is increased, and is proportional to $\lambda \sim W$ and $\lambda \sim W^{1/2}t^{1/2}T^{-1/4}$, respectively, where $W$ is the width of the ribbon, $t$ the thickness, and $T$ is the normalized applied tension.

Experiments were performed with mylar and cellulose acetate sheets which have a linear elastic response for strains less than 2\% (see Supplementary Documentation.) Above the elastic limit, mylar deforms plastically but does not rupture. Where as, cellulose acetate is a quasi-brittle material which leaves a contrasting mark when deformed just above the elastic limit and can then rupture.   Young's modulus $E$ and Poisson ratio $\nu$ for Mylar are $E \simeq 3.4$\,GPa and $\nu = 0.4 \pm 0.05$, and for cellulose acetate $E\simeq 2.2$\,GPa and $\nu = 0.35\pm 0.05$. Ribbons with thickness $t = 75$, $125$, and $256\,\mu$m, width $W$ in the range $10-30$\,mm, and length $L=100-300$\,mm are used. The ends of the ribbon are clamped and  stretched by applying a constant force $F$, and then twisted around its long axis by a prescribed twist angle $\theta$. Therefore, the experimental control parameters are the normalized tension $T={F}/{Etw}$ and normalized twist angle $\eta = \theta W/L$. Accordingly, one can define a confinement parameter $\alpha = \eta^2/T$ which is the ratio of a geometrical strain over a mechanical strain~\cite{King2012}.  

\begin{figure}
\centering
\includegraphics[width = 8cm]{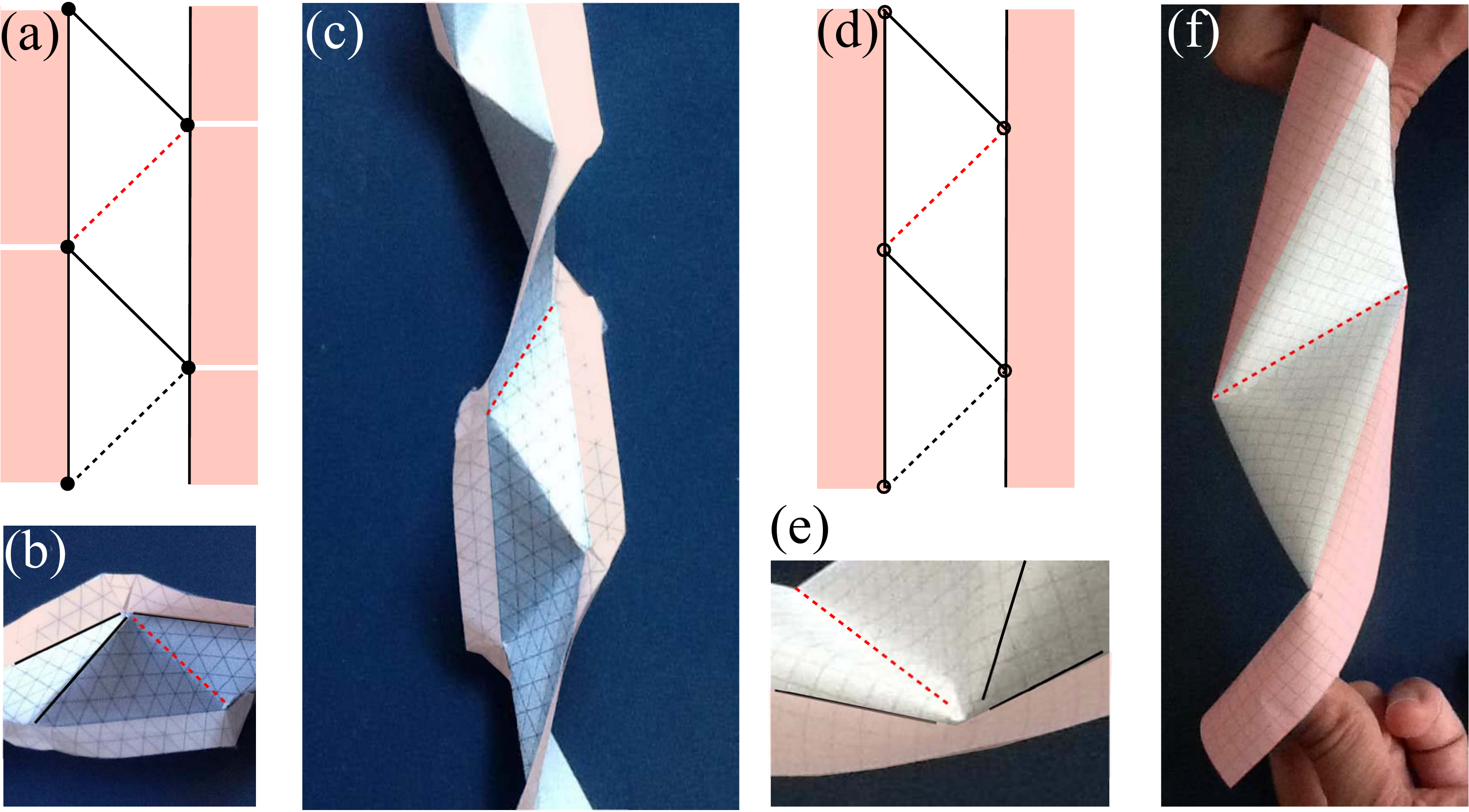}
\caption{A paper model of a ribbon with e-cones (a-c) and d-cones (d-f) which is constructed by hand on a triangular lattice. (a) A ribbon with e-cones (black filled circles) obtained by making cuts (white lines) and inserting $\pi/6$ angled wedges in the lateral flaps (pink/gray areas). (b) A close-up of an e-cone and associated ridges. (c) An e-cone decorated helicoid at equilibrium with flat triangular regions connected by ridges. (d) A ribbon with d-cones (black open circles) and the two lateral flaps. (e) Close-up of a d-cone with lines defining ridges and the inner edges of the flaps. (f) A twisted helicoid with d-cones with flat triangular regions connected by stretched ridges.}
\label{fig2}
\end{figure}
In order to measure its morphology, the ribbon is scanned as a function of applied twist and the surface identified by using a threshold contrast for the absorbed x-rays (see Supplementary Documentation.) As shown in Fig.~\ref{fig1}, the longitudinal and transverse coordinates along the ribbon surface normalized by $W$ are denoted by $s$ and $r$, respectively.  The Gaussian curvature $K$ and mean curvature $H$ are then obtained by locally fitting the surface with a quadratic function. A reconstructed shape of the ribbon along with $K$ and $H$ superimposed on a single ridge is shown in Fig.~\ref{fig1}(a) and Fig.~\ref{fig1}(b), respectively.  One observes that regions with higher $K$ are confined to the vertices, and those with higher $H$ are distributed along the ridge as well as focused near the vertices. Further, the peak values of the curvatures do not coincide spatially. Thus, the actual route toward localization and defect formation can be quantified by the evolution of $H$ and $K$ using $\eta$, or equivalently with $\alpha$, for a fixed $T$.

\begin{figure}
\centering
\includegraphics[width =8cm]{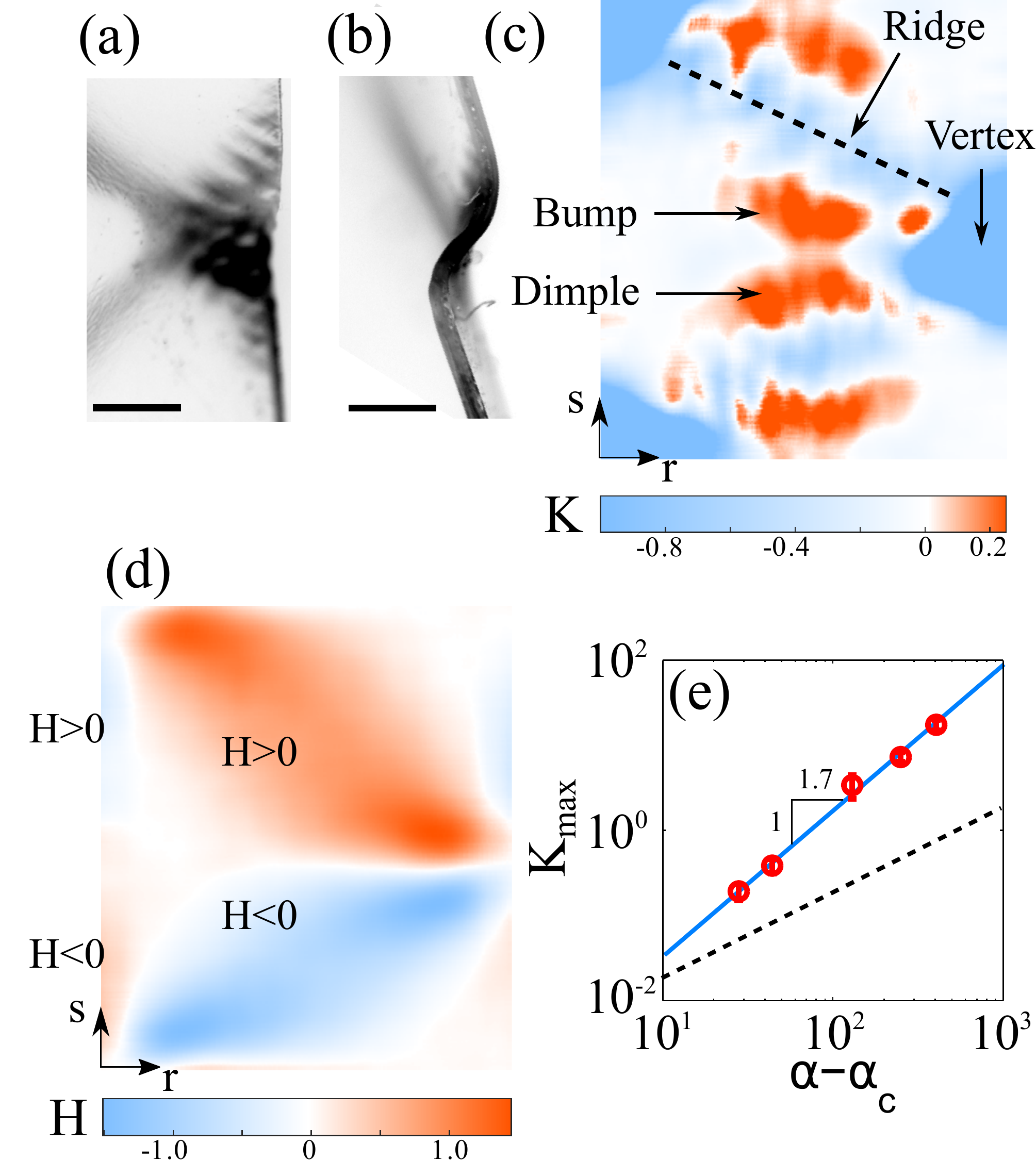}
\caption{(a) Front view and (b) side view of the plastic deformations located near the vertex in a cellulose acetate sheet ($t/W = 7.4 \times 10^{-3}$, $T = 2.1\times 10^{-4}$ and $\alpha = 3700$) indicating that the strain is mainly localized in a small triangular wedge along the ribbon edges and along the stretched ridges. The side view shows a change of curvature along the edge when passing over the defect. Scale bar is 1mm. (c) Gaussian curvature of a ribbon (Mylar, $t/W = 8.5 \times 10^{-3}$, $T = 1.2 \times 10^{-3}$) for $\alpha = 430$ shows stretched ridges with $K<0$ connecting e-cones. In-between the ridges, bumps and dimples ($K>0$) under low compression  are observed. Data for $K > -1$ is shown for better visualization. (d) Mean curvature of the same ribbon as in (c) showing two ridges of positive (red/light gray) and negative (blue/gray) curvature. 
(e) The maximum $K_{max}$ of the Gaussian curvature on the defects increases as a power with $\alpha$
Note $K_{max} \gg -\eta^2$ (dashed black line).}
\label{fig3}
\end{figure}
\begin{figure}
\centering
\includegraphics[width = 8.5cm]{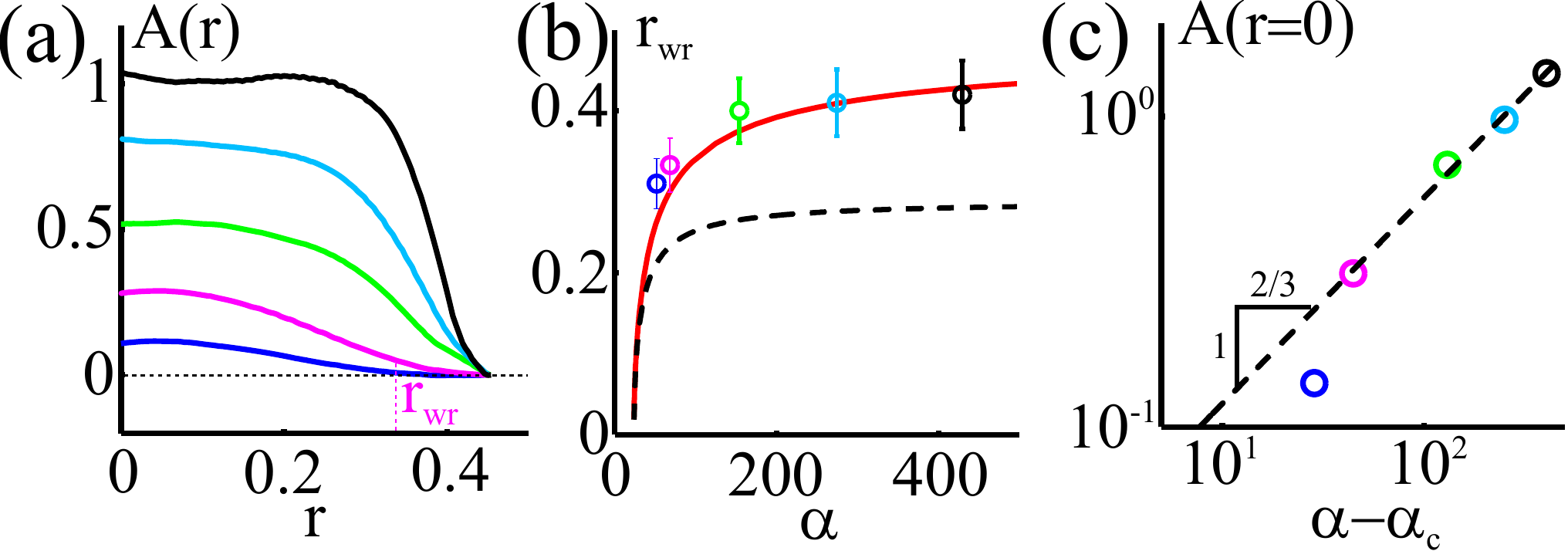}
\caption{(a) Profile of mean curvature amplitude $A(r) = \sqrt{\langle H^2 \rangle _s}$ for an increasing $\alpha$ showing the growth of amplitude and width of the wrinkles. (b) Evolution of the width of the wrinkles $r_{wr}$ with $\alpha$ is in agreement with the form predicted using a far from threshold analysis (solid red line). For comparison, the prediction from a near threshold analysis is plotted (dashed black line). (c) Evolution of the amplitude $A(r = 0)$ with $\alpha$ showing a sublinear dependence with an exponent $0.65(\approx 2/3)$ measured for large $\alpha$.}
\label{fig4}
\end{figure}

As is well known, a twisted ribbon has a helicoid shape with $H \approx 0$ and  $K = -\eta^2$ for small enough $\alpha$ at finite $T$,  and wrinkles just above $\alpha > \alpha_c \approx 24$  due to the development of compressive stress around the centerline~\cite{Green1937,Coman2008,Chopin2013}. Accordingly, the map of $H(r,s)$ and $K(r,s)$ is shown  for $\alpha > \alpha_c$ in Fig.~\ref{fig1}(c) and Fig.~\ref{fig1}(d), respectively. We observe a continuous transition from smooth wrinkles to sharp ridges along with an increase in the overall curvature by an order of magnitude as $\alpha$ is increased from 55 to 437. Further, while the wrinkles are initially confined to the center of the ribbon,  they are observed to grow in width as $\alpha$ is increased, before a symmetry breaking occurs when the region with larger $H$ tilts with alternating angles. In contrast, the regions with larger $|K|$ progressively reduces in size at the expense of those with smaller $|K|$. For $\alpha \gg \alpha_c$, $H$ is localized along ridges with alternating tilt angles, while $K$ is localized in small regions which are self-organized in a triangular lattice with a spacing similar to the wavelength at threshold. As we discuss next, these localized regions can be thought of as point-like defects which can be modeled as conical singularities with a geometry which can be assigned using the measured curvature. 

To understand the nature of these singularities, we first examine a model of conical singularities prescribed at the vertices of a triangular lattice which are located at a finite distance from the edge (see Fig.~\ref{fig2}.) In practice, negative and positive disclinations are obtained by adding or removing a wedge-shaped sector at the edge of the ribbon, while 
d-cones are created by forcing a point on the sheet -- corresponding to the vertex -- into a small rim~\cite{Cerda1998}. We observe that the buckled shape is qualitatively different from that of an isolated singularity because of the interaction between defects. In Fig.~\ref{fig2}(a-c), we show that a ribbon 
with embedded e-cones leads to a triangular faceted helicoid that we name an e-helicoid. The e-helicoid is reminiscent of the wavy edges of torn polyethene sheets and leaves induced by plastic flow and growth~\cite{Sharon2004}, except that in our case, the Gaussian curvature is localized due to the underlying non-planar configuration of the ribbon. We further found that a helicoid structure could not be constructed using positive disclinations. However, d-cones organized in the same triangular lattice (see Fig.~\ref{fig2}(d)) lead to a faceted helicoid upon twisting  with very similar triangular facets connected by stretched ridges (see Fig.~\ref{fig2}d-f). 
This is a significant result as it is highly non-trivial to find isometric configurations of elastic sheets theoretically or numerically with given boundary conditions. These ordered developable shapes, that we call d-helicoids, offer an intermediate step toward understanding more complicated, fully disordered crumpled sheets which are also singular developable shapes with vertices and ridges.

In order to compare to these models, we display the front and side views of a scar formed on a twisted ribbon due to plastic deformations located at the core of a singularity in Fig.~\ref{fig3}(a) and Fig.~\ref{fig3}(b). Unlike d-cones which form a parabolic scar~\cite{Cerda1998}, we observe a triangle-shaped plastic deformation near the edge. In this region, we note that the metric of the surface is similar to the paper model of an e-helicoid shown in Fig.~\ref{fig2}(c). Further, these characteristic features can be compared with the measured Gaussian and mean curvatures shown in Fig.~\ref{fig3}(c) and (d), respectively, where the curvatures are averaged over several ridges to improve the signal to noise ratio. As shown in Fig.~\ref{fig3}(e), the core of the defects have a strong negative Gaussian curvature which increases with $\alpha$ as $K_{max}\approx K_0 (\alpha - \alpha_c)^{1.7}$, with $K_0 = 6.7 \times 10^{-4}$. We observe that the cores of these singularities are connected by ridges with large $H$ with alternating signs and relatively small $K$ which shows stretching between interacting e-cones. Further, in between ridges, we identify regions with positive $K$ and alternating $H$ indicating bumps and dimples due to compression. As noted in the discussion of the model of an e-helicoid in the previous paragraph and Fig.~\ref{fig3}(b), lateral flaps located near the edges of the ribbon ($r \approx \pm 1/2$) are developable ($K \approx 0$) but with small curvature of alternating signs (see Fig.~\ref{fig3}(d)). Based on these measurements and comparisons, we conclude that a stretched twisted ribbon spontaneously forms e-cones, rather than d-cones, in the far from threshold regime.

We next 
focus on the
compressed section in the central part of the ribbon.
Taking advantage of the translation invariance along the $s$ direction, we introduce the characteristic amplitude $A(r) = \sqrt{\langle H^2 \rangle _s}$ of the wrinkled section averaged along $s$. With this definition, $A(r)$ is identically zero and positive for developable and helicoid states, respectively. In Fig.~\ref{fig4}(a), we plot $A(r)$ for increasing $\alpha$. Because the transition from the wrinkled section to the stretched flaps appears continuous,  the wrinkled section is defined in practice as the region where $A(r) > A(0)/10$ corresponding to $r<r_{wr}$. As shown in Fig.~\ref{fig4}(b), the evolution of $r_{wr}$ with $\alpha$ compares well with the prediction from a far from threshold analysis which assumes that the formation of the wrinkles relaxes the stresses leading to a compression free central part and tensile flaps~\cite{Davidovitch2011,Chopin2015}.  By contrast, a linear perturbation analysis does not capture the evolution of $r_{wr}$ (dashed black line). This is because assuming that the wrinkles do not relax the stress leads to unphysically large compressive stresses especially in the case of $\alpha \gg \alpha_c$. Further, the evolution of the wrinkling amplitude $A(0)$ evaluated at $r=0$ is observed to increase sub-linearly for large $\alpha$, consistent with $A(0) \sim (\alpha-\alpha_c)^{2/3}$. However, the origin of this scaling remains unclear.

\begin{figure}
\centering
\includegraphics[width = 6.5cm]{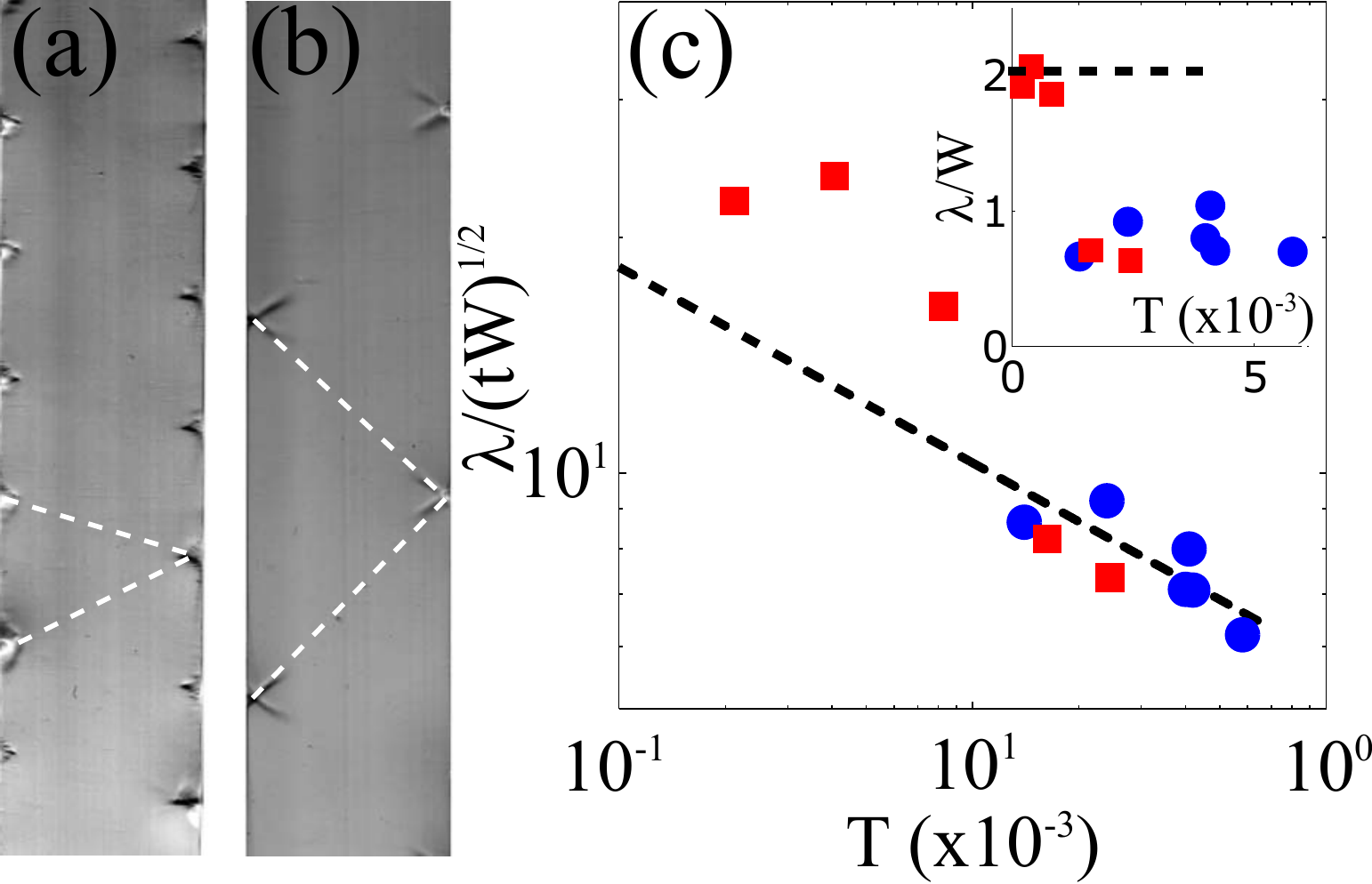}
\caption{Twisted ribbons (cellulose acetate) with  $T = 2.5 \times 10^{-3}$ (a) and $T = 2.1 \times 10^{-4}$ (b) showing plastics deformations at the edges with a spacing which decreases with tension. The dashed lines represent location of the ridges which connect the e-cone singularities to guide the eye. (c) Normalized spacing between defects $\lambda / \sqrt{tW}$ decreasing with the tension (for $T > 10^{-3}$) using Mylar (blue circles) and cellulose acetate (red squares). The prediction near threshold $C \lambda_{NT} / \sqrt{tW}$  (black dashed line) provides a reasonable fit to the data using the adjusting parameter $C = 0.9$. 
The wavelength is observed to saturate at $\lambda/W \approx 2$ 
for $T < 10^{-3}$ (see inset).}
\label{fig5}
\end{figure}

Finally, we analyze the lattice spacing between defects $\lambda$.  As shown in Fig.~\ref{fig5}(a,b), we find that $\lambda$ decreases with $T$ and that there is little dependence with $\alpha$. We verified that the defects are not trapped in one location due to plasticity by decreasing $T$ under fixed twist angle. 
Two regimes for the defect spacing are identified (see Fig.~\ref{fig5}(c)). At relatively large tension ($T>10^{-3}$) corresponding to the elastic regime, $\lambda$ is observed to clearly decrease with tension. Because the scaling for $\lambda$ is not  available in the far from threshold regime,  we compare the experimental data with the prediction from a linear stability analysis 
which gives  $\lambda_{NT} = 1.71(1-\nu^2)^{-1} \sqrt{t/W}T^{-1/4}$~\cite{Coman2008}. We find that the data can be well described by $\lambda_{fit} = C \times \lambda_{NT}$, where $C=0.9$ is an adjustable non-dimensional prefactor 
close to one. This indicates that the scaling of $\lambda$ does not change significantly far from threshold, as opposed to the scaling of $r_{wr}$ shown in Fig.~\ref{fig4}(b) which is not described by near threshold analysis.  Similar distinction has been noted recently as well in the wrinkling of an elastic sheet floating on a liquid drop~\cite{King2012}. However, at lower tension, a plateau is observed for $\lambda \sim 2 W$, with no observable dependence on $T$ (Fig.~\ref{fig5}(c), inset) indicating a lattice spacing essentially controlled by the geometry.

In conclusion, we find that a thin extensible ribbon upon twisting exhibits a continuous transition from smooth wrinkles to sharp ridges. We show with physical models that faceted helicoids can be constructed using ridges that connect either d-cones or negatively signed disclinations i.e. e-cones that are located in a periodic triangular lattice near the edges of the ribbon. Our measurements clearly demonstrate that the singularities in extensible ribbons are e-cones, and we thus call the resulting structure an e-helicoid. The longitudinal stretch of the ribbon is found to control the growth of the e-cones and their spatial organization with a lattice spacing depending on ribbon geometry and elasticity. Hence, we demonstrate that the stretched twisted ribbon has a dual nature as it shows properties intermediate between a torn polyethene sheet where strain induced plasticity leads to  edges with negative Gaussian curvature and a crumpled paper where the inextensibility condition produces shapes with flat facets and localized defects. Finally, we suggest that the characteristic features of interacting e-cones found here may be applied to other thin film systems. They may help understand extensible sheets under other conditions which result in changes to the metric, examples of which include torn thin films, shells with cracks, film growth, and even crumpled sheets. 

We thank Marcelo Dias, Vincent D{\'e}mery, Martin Mueller, Basile Audoly, Benjamin Davidovitch, Pascal Damman and Mokhtar Adda-Bedia for stimulating discussions. This work was supported by National Science Foundation under grants DMR-0959066 and DMR-1508186. 

\bibliography{econearxiv}

\end{document}